# Analytical tools for single-molecule fluorescence imaging *in cellulo*

M.C. Leake [a]

Recent technological advances in cutting-edge ultrasensitive fluorescence microscopy have allowed single-molecule imaging experiments in living cells across all three domains of life to become commonplace. Single-molecule live-cell data is typically obtained in a low signal-to-noise ratio (SNR) regime sometimes only marginally in excess of 1, in which a combination of detector shot noise, sub-optimal probe photophysics, native cell autofluorescence and intrinsically underlying stochasticity of molecules result in highly noisy datasets for which underlying true molecular behaviour is non-trivial to discern. The ability to elucidate real molecular phenomena is essential in relating experimental single-molecule observations to both the biological system under study as well as offering insight into the fine details of the physical and chemical environments of the living cell. To confront this problem of faithful signal extraction and analysis in a noise-dominated regime, the 'needle in a haystack' challenge, such experiments benefit enormously from a suite of objective, automated, high-throughput analysis tools that can home in on the underlying 'molecular signature' and generate meaningful statistics across a large population of individual cells and molecules. Here, I discuss the development and application of several analytical methods applied to real case studies, including objective methods of segmenting cellular images from light microscopy data, tools to robustly localize and track single fluorescently-labelled molecules, algorithms to objectively interpret molecular mobility, analysis protocols to reliably estimate molecular stoichiometry and turnover, and methods to objectively render distributions of molecular parameters.

## Introduction

Ernest Rutherford, the father of modern nuclear physics who was awarded the 1908 Novel Prize in Chemistry, is attributed to have commented prior to this that all science was either 'physics or stamp collecting'. What Rutherford lacked at this time was a clear insight into the enormous future extent of modern interdisciplinary research between the physical and life sciences that would emerge in the century to follow and beyond; there is currently now a furious energy of cutting-edge research activity at this interface which only the truly naïve would label as 'stamp collecting'. Experimental single-molecule biology research in its most modern and exciting form essentially combines state-of-the-art 'wetlab' approaches from the life sciences with cutting-edge technology and intellectual insight and rigor from the physical, mathematical, computational and engineering sciences, in challenging assays that aim to preserve the underlying physiological context, namely to perform experiments on living cells. *In vivo* single-molecule approaches, arguably better termed '*in cellulo*' approaches if one is observing single individual cells to discriminate from investigations on multicellular organisms (excepting single-celled organisms such as bacteria which can be described in both contexts), add significant insight not only into the native biochemistry of the living cell, but also the important physical chemistry and chemical physics of the cellular environment – the ionic strength, pH, viscosity, microrheological features, phase transition behaviour, osmolarity, as well as a breadth of information concerning free energy landscapes of observed molecular components.[1-4]

The primary importance of single-molecule live-cell data is due to often highly heterogeneous behaviour in such physical parameters both as a function of localization in the cell and of history-dependent effects in the cell cycle. Namely, there are both spatial and temporal dependencies in the internal environment of cells, which therefore require not only a functioning cell as an experimental specimen but also a probe at the molecular length scale of the nanometre to offer a level of spatial precision sufficiently high to probe the highly local fluctuations in physical chemistry that can potentially occur on the nanoscale – even the most basic of living cells from the prokaryotic domain, such as bacteria, are far more than just static bags of chemicals, but rather are dynamic structures which have distinctly defined molecular architectures at the sub-cellular level affecting the physical chemistry of the internal cellular environment.[5]

Single-molecule data is intrinsically stochastic in nature, which can lead to potential ambiguity in interpretation in the absence of robust analytical tools. Fluorescence imaging data at the single-molecule level from living cells also suffers from further detection challenges since they are compounded with the effects from a variety of sources of noise, including shot noise at the level of the camera detectors, photophysical spectral shifts and broadening of the fluorescent reporter dyes, native cellular autofluorescence and out-of-focus contributions of fluorophores, in addition to nearest-neighbour issues in situations of high molecular concentrations making definitive continuous detection of the same fluorescently-labelled molecule over extended time scales challenging.

Furthermore, the majority of *in cellulo* single-molecule fluorescence imaging utilizes fluorescent protein labelling at the level of fusion of the encoding DNA[5,6] and these naturally occurring dye molecules in general have poor absorption cross-sections and low photostability compared to synthetic organic dyes, manifest in comparatively low fluorophore brightnesses and short photoactive lifetimes. In essence, typical single-molecule fluorescence imaging datasets have a poor equivalent signal-to-noise ratio (SNR) in reference to the typical pixel intensities registered on camera detectors, and are often relatively short in duration, for example consisting of 10 or less consecutive image frames.[7,8]

However, it is also not uncommon to acquire significant quantities of such datasets during experimental runs, essential in constructing the underlying probability distribution of heterogeneous and stochastic molecular behaviour. The need for objective, robust, automated high-throughput and computationally efficient analysis tools is imperative to determine the underlying molecular properties that are markers of biochemical, physical chemical and chemical physics features of the internal cellular environment.

The use of single particle tracking (SPT) techniques to investigate biological processes is a well-characterized and popular approach, especially so in the context of monitoring fluorescent reporters, either single molecules or nanoscale particles such as fluorescent beads or quantum dots, that, in the case of fluorescent protein molecules, can be tagged at the level of the encoding DNA to protein molecules of interest in a living cell.[9-15] These methods have been applied to, for example, estimates of the diffusion coefficient of individual protein complexes in the cellular membrane[16] and investigating the molecular stoichiometry and turnover of molecular machines *in vivo*.[17,18] While green fluorescent protein (GFP) and its variants offer enormous potential for increasing our understanding of biological processes in large data sets, automatic tools with efficient techniques for analyzing time-lapse microscopy images from real dynamic cell processes are essential for objective and systematic quantization.[19-24]

**Analytical tools for interrogating single-molecule data *in cellulo***

**1. Automated segmentation of cellular images from light microscopy**

The dependence of spatial localization of physical and chemical properties in living cells necessitates objective methods to determine the true spatial extent of cells from light microscopy images. However, light microscopy images of cells, even when acquired using contrast enhancement methods such as phase contract or differential interference contract, result in a blurring of the true cellular boundary due to convolution by the point spread function (PSF) of the imaging system, necessitating software tools to determine the precise position of the cellular boundary.

Many recent image processing techniques have been proposed to determine the precise boundaries of cells with the external environment. Image segmentation approaches can extract the cellular characteristics of size and shape. Pixel intensity thresholding methods are useful and fast if the cell body intensity is homogenous and easily distinguished from the non-cellular background, but typically produce jagged edges that do not accurately represent the smooth cellular boundary.[25] There exist more recent advanced approaches that are more robust, including Region Growing methods,[26] often generating better results when the pixel intensity in foreground cell objects of the microscope image are homogenous, but arguably are of more limited use for noisy experimental fluorescence imaging data of the type typical for imaging of fluorescent protein constructs. Watershed methods utilize 'flooding' strategies along the intensity baseline, being particularly sensitive to distinct gradient changes at object boundaries; varied algorithms of this type have proved to be effective and have been used in several successful software applications, including CellProfiler (http://www.cellprofiler.org). Level Set algorithms offer a good option in cellular image segmentation by creating higher dimensional space from the multiple image parameter set typically used, and find optimal solutions to this using multi-parameter minimization algorithms; the quality of output that results is often high when the input information utilizes multiple experimental data from intensity, cell boundary position and prior cell shape knowledge.[27]

Recent image segmentation methods developed in my own laboratory have obtained more reliable results through the *Maximum a Posteriori* (MAP) method[28-31] by minimizing an objective energy function obtained from the raw image data.[32] The foreground image intensity and geometrical boundary of each cell can be modelled, and the compactness and smoothness can be estimated. Simulated Annealing (SA)[33,34] can help estimate parameters, and our approach is to apply this method to generate a sub-pixel precision of the cell boundary position. The method exploits a generative model to determine the shape based only upon pixel intensity, so it is comparatively robust.

A significant number of cell types have well-defined shapes and sizes, seen for example in the rod-like *Escherichia coli* bacteria used in a number of recent *in cellulo* fluorescence imaging studies to investigate both internal cell process and those integrated in the cell membrane. In such cases of investigating cell membrane processes the cell shape may be characterized relatively easily allowing global Cartesian coordinates on the camera detector plane to be transformed onto the local membrane surface coordinates relevant to single imaged cell. For biological membrane protein systems, sub-pixel changes in position may occur over a time scale of milliseconds. These can potentially be quantified accurately using a local coordinate representation in combination with superresolution localization imaging for even nanoscale movements of fluorescent protein reporters, for example detecting tiny gene expression bursts at the single-molecule level the appearance of membrane protein reporters.[35]

There is common agreement on a canonical shape for *E. coli* bacteria, namely that of a cylindrical body capped by hemispheres, with a typical diameter width of ~1 μm and end-to-end length in the range ~2-5 μm, depending primarily on stage in the cell cycle.[7,8,16-18] Here, we define a local coordinate system from each bacterium's position, orientation, length and width, as estimated from the non-fluorescence brightfield imaging data. Through observations using light microscopy we can characterize the size and shape of each detected *E. coli* cell individually, with its characteristic 'stubby' object shape[32] shown in Fig. 1.

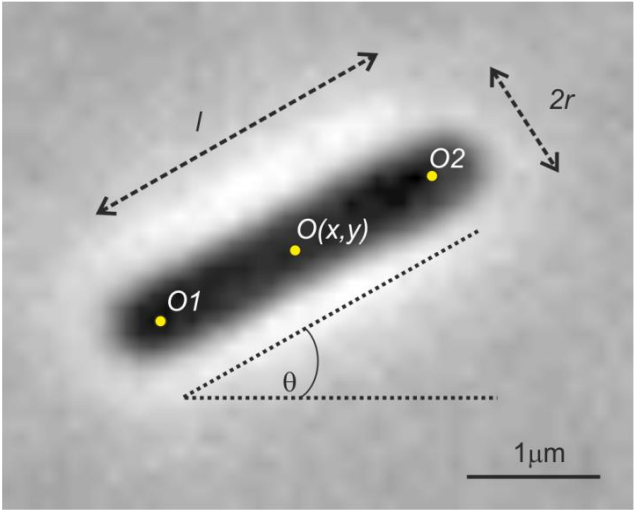

Fig. 1 Brightfield microscopy image of a single *E. coli* cell (greyscale) modelled as cylinder with hemispherical caps, with parameter set to reconstruct cell body perimeter projected onto the two-dimensional plane of the camera detector.

Each *E. coli* cell shape can be fitted to an observed cell object image by estimating parameters from within a sensibly defined region of parameter space. The match between cell object and characteristic shape quantifies the particular manner in which the parameter set is chosen by estimating optimal parameter sets from an object energy function.

For constructing a local coordinate system relevant to single *E. coli* cells, four parameters characterize the two-dimensional cell boundary: (i) a local origin – $O\{x,y\}$ unique to each cell with respect to the global (camera) coordinate system; (ii) long axis orientation of the cell with *x*-axis of the camera $\{\Theta\}$; (iii) length – $\{l\}$ which is cylinder long axis length; (iv) radius – $\{r\}$ which is the radius of hemispherical cap and equal to half the cylinder diameter. To estimate these properties we adopted a stochastic framework within a multivariate statistical model.

The parameter set of $\{x,y,\Theta,l,r\}$ allows us to compute several other relevant attributes, e.g. total length of the cell body parallel to the long-axis as the sum of $l+2 \cdot r$, the width is $2 \times r$, the centre position of the two hemispherical caps $O_1$ and $O_2$ are computed by $\{x \pm l/2 \cdot \cos\Theta, y \pm l/2 \cdot \sin\Theta\}$. The origin of the local coordinate system can be defined as either $O$ or $O_1$ on the different applications depending on the specific application desired.

For the microscopy images, we need to consider the PSF, which accounts for the blurred cell boundary appearance and resultant ambiguity over the precise boundary location. We introduce a variable $\{f\}$ to indicate the PSF in the above defined parameter space $\Omega$. This parameter is not required to directly determine the standard shape, but it is important to deconvolve the raw blurred cell image in the pattern matching process, so it is a latent variable in $\Omega = \{x,y,\Theta,l,r,f\}$. The task of determining the local coordinate system of a given shape representation is an optimal parameter estimation problem $\Omega^* = \{x^*,y^*,\Theta^*,l^*,r^*,f^*\}$.

A light microscopy live-cell image can be modelled as a 'foreground' object of interest, on top of an image 'background'. The foreground object can be represented as a two-dimensional matrix which can in turn be rendered into a one-dimensional vector to be processed. Let *I* be the observed two-dimensional matrix image with size $N \times N$ pixels, resulting from an uncontaminated, noise-free foreground image $I^*$ plus a zero-mean Gaussian noise image *C*, which can be expressed as $I = I^* + C$, where $C \sim N(0,\sigma_c)$, with $\sigma_c$ image noise standard deviation. Similarly, the parameter set representing a given image can be represented by the idealized optimal parameter set with the addition of some parameter perturbation. Considering the parameter vector space $\Omega$ with size M, we model each estimated parameter element as an individual random variable consisting of real value $\Omega^*$ plus zero mean Gaussian distribution as $\Omega = \Omega^* + D$, where *D* is zero-mean Gaussian distribution $N(0,\sigma_d)$ which reports the noise associated with the given parameter element. The conditional probability distribution of these estimated parameter variables can be formulated as:

$$P(\Omega / \{\Omega_i\}, \{\sigma_{Di}\}) = \prod_{i=1}^{M} p(\Omega_i / \Omega_i, \sigma_{Di}) \quad (1)$$

Using Bayesian inference, $p(\Omega|I) = p(I|\Omega) \times p(\Omega)/p(I)$. A point estimate $\Omega^*$ may be determined by maximizing the conditional distribution $p(\Omega|I)$. The main challenge lies in estimating the unknown shape coefficients $\Omega^*$ obtained only from the original image $I^*$. We use MAP to define a statistical model allowing us to maximize $p(\Omega|I)$. We can transform the probability distribution to rewrite Equation 1, and $\Omega^*$ can be obtained as follows:

$$\Omega^* = \arg\max_{\varphi \in \Omega^T} \{\ln(p(\Omega/I))\} = \arg\max_{\varphi \in \Omega^T} \ln\left\{\frac{p(I/\Omega) \cdot p(\Omega)}{p(I)}\right\}$$
$$= \arg\max_{\varphi \in \Omega^T} [\ln p(I/\Omega) + \ln p(\Omega) - \ln p(I)]$$
$$\propto \arg\max_{\varphi \in \Omega^T} [\ln p(I/\Omega) + \ln p(\Omega)] \quad (2)$$

Here, *I* is the recorded microscopy image on the camera detector, $\varphi$ is the candidate set for estimating parameters and $\Omega^T$ is a uniformly spaced set of points in the parameter space. In this formulation, we assume that the experimental images *I* are consistent with the model $\Omega$ with prior distribution $p(\Omega)$.

The maximum a posteriori estimate for a given set of the shape model parameters is dependent on the sum of both $ln(p(I|\Omega))$ and $ln(p(\Omega))$. The $ln(p(I|\Omega))$ component is a measure of the likelihood of the original image given parameters $\Omega$. We characterize the variation in this model space by a Gaussian distribution $N(0,\sigma^2)$.

We use the spatial intensity distribution inside the boundary of each individual *E. coli* cell as the equivalent likelihood function, and the precise detected cell boundary edge of each cell image provides additional information for matching a well-defined template shape to modify the bias from the maximum likelihood formulation. In noisy, complex live-cell microscopy images there may be non-trivial ambiguity as to where this precise cell image edge is located. Previous studies in image analysis have shown that the cell image edge may be modelled using a monotonically decreasing intensity function[36] on top of a Markov random field (MRF),[37,38] consistent with empirical observations from raw images. An MRF provides a method to model the joint probability distributions of the image sites in terms of local spatial correlations, which can be expressed by the energy function $U(I)$, which is a measure of the spatial connection of pixels across the image, known as the 'clique' $c$. The energy from $c$ is defined as $E(p_o, p_i) = \beta$, where $p_0$ is the pixel of interest, and neighbouring pixels are signified as $p_i$ $(i=1,..,n)$, to construct pair-wise clique values with $n$ elements in total. $\beta$ is a positive number describing the strength. For non-neighbouring pixel elements, the clique energy is assumed to be zero.

The prior distribution of shape parameters can be regarded as a penalty to control the bias from the intensity likelihood method, which can be treated as a regularized selection in MAP. Using knowledge from a prior distribution of image shape can generate an improved estimate of the cell image edge than considering just likelihood component alone. Empirical microscopy observations indicate that in the case of *E. coli* cells their width is relatively constant to within ~10%.[7,8] Thus we can give a sharp prior statistical distribution for cell width and use the uniform function as prior distribution for the other variables. Starting from the MAP-based formulation, our method combines three key features from the image object for parameter estimation: pixel intensity $p(\Omega|I)$, edge penalty $p(I)$ and prior distribution $p(\Omega)$.

The SA method is a standard optimization tool, and has proved to be an effective method for MAP optimization in particular. This problem can be resolved by using an optimal searching strategy to get the best candidate, which will match the observed object accurately by generating a standard shape. We use the SA algorithm to find the global optimal resolution $\Omega^*$.

To test the performance of our cell shape reconstruction algorithm we simulated images containing bacterial cells randomly distributed over the sample surface with their long axes assumed parallel to the surface, with cells having variable end-to-end lengths and diameters over a physically sensible range typical of real bacteria. Background noise levels were increased systematically to give a signal-to-noise ratio (SNR) from 2 to 5. Here, SNR is defined as $SNR = S/\sqrt{S+N}$, where $S$ and $N$ are the mean foreground signal pixel intensity, and mean background noise pixel intensity respectively. Because the bacterium observed in light microscopy has an indefinite boundary from the background, the estimated shape may not fit the boundary of object accurately in the case of low SNR.

Simulated datasets were analyzed using our bespoke software and the real values recorded before the convolution of PSF and pixel sub-sampling (Fig. 2).

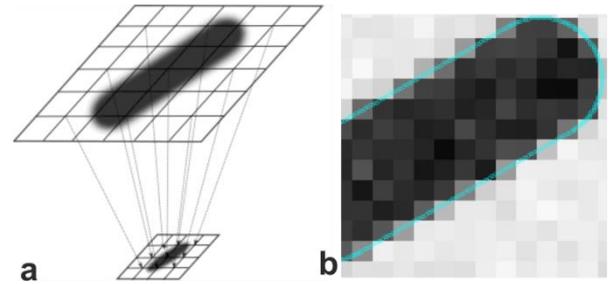

Fig. 2 Sub-sampling technique from (a) large scale image to small scale image is used for simulating a bacterial cell image obtained from an EMCCD camera detector; (b) The imaging system's point spread function is used in the deconvolution of cell boundary from the foreground; (c) Simulated *E. coli* with exact boundary (cyan) controlled to sub-pixel accuracy.

To evaluate the performance of this shape reconstruction approach, we compared the results from several reference methods. Of the conventional segmentation tools currently available we found that the marker-controlled watershed transformation segmentation algorithm,[39] which relies on defining quantitative measurement of the intensity gradient transform, gave the most promising results using our brightfield experimental data of live *E. coli* cells, which could in many cases be used to determine the location of the cellular boundary to a sub-pixel precision.

The basic watershed transformation works in the following way. Firstly, any greyscale image can be modelled as a topographical surface. If we then flood this surface from its lowest point, and prevent any water combining from different sources, then 'watershed lines' between 'catchment basins' can be defined marking out the lines of image segmentation. An improvement to this involves the uses of pre-selected markers from which to initiative the flooding, which minimizes the risk of over-segmentation in the image.

In Fig. 3 we compare the results of simulated noisy brightfield cell images using our generative MAP shape reconstruction algorithm with the marker-controlled watershed method.

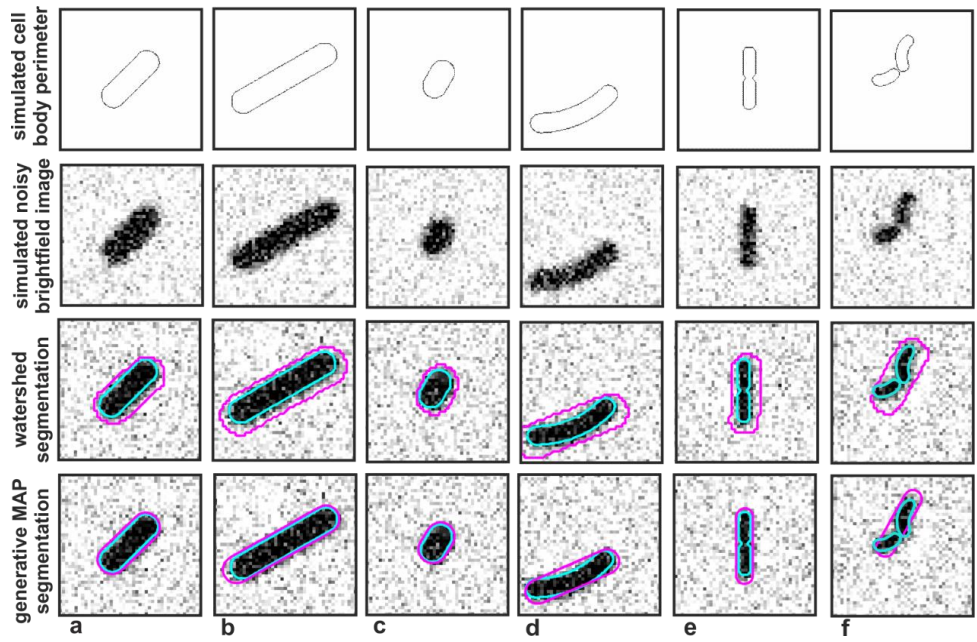

Fig. 3 Comparison of performance of cell body segmentation methods using the watershed method or our generative MAP method for different categories of cell body shape for (a) 'half fat', (b) 'elongated thin', (c) 'short fat', (d) 'banana', (e) 'lotus', (f) 'cluster fat' categories; in lower two panels simulated cell body perimeter (cyan) and the predicated perimeter from the respective segmentation algorithm (magenta) are indicated.

The figure shows a series of example shapes with the simulated cell boundary which having several varying length and width combinations. The most commonly observed category of *E. coli* cell shape, the standard rod type shape with an *l/r* value of ~2, we denote as the 'half fat' shape. However, sometimes bacteria are not distributed uniformly flatly over the surface of the microscope coverslip parallel to the cells' long axis, in addition cells will elongate during the cell cycle prior to cell division and may have slightly curved appearances, these factors resulting in some cells appearing closely together as a cluster of bacteria, sometimes with cell bodies projecting above the coverslip surface, so the inferred cell length from the two-dimensional image may be artefactually low. To characterize these effects we could also simulate such clusters/overlaid cells, as well as elongated and curved variants. All of our synthetic shapes consist of six underlying categories: {'half fat', 'elongated fat', 'short fat', 'curved fat', 'lotus fat', 'cluster fat'}, shown in Fig. 3 with these simulated examples all having the same SNR of 6.

To estimate the objective measure of the accuracy in the reconstruction method we calculated the residual mean-square error (MSE) to compare the simulated with the measured parameters. This indicates that our MAP approach has an intrinsic accuracy in the absence of noise of MSE ~0.01 pixel. The MSE in the presence of noise, using an SNR of 6 typical for our real brightfield microscopy images, is a measure of the effectiveness of the shape reconstruction method under experimental conditions, indicating that our generative MAP segmentation algorithm generates an output typically within 1-2 pixels of the true cell boundary position, superior to the watershed method by a factor typically greater than 2.

**2. Pinpointing fluorescently-labelled molecules**

The first step in pinpointing where exactly a fluorescently-labelled molecule is in a living cell is to attempt to restore the uncontaminated noise-free image, since a key practical challenge of dynamic fluorescence microscopy *in cellulo* is often extremely noisy data due to several reasons, e.g. dark states of the fluorescent reporter tag, especially blinking of quantum dots and fluorescent proteins, background fluorescence due to parental autofluorescence from cells, shot noise for the camera pixel readout in the background, as well as Poisson signal noise and pixilation noise (the latter is essentially due to uncertainty as to where fluorescence photons originated from if detected within a single pixel whose length scale when projected onto the level of the image is typically 50-150 nm).

Microscopy image data consists typically of background fluorescence and cellular structures with fluorescence levels of intensity close to the peak intensity emitted from fluorescent particles. The image of a fluorescent particle being tracked, with intensity consists of a real fluorescence photon component associated with each particle, plus possible noises contributing the sum of a final observed fluorescence peak.[40]

For any observed fluorescent particle in the image, the number of detected fluorescence emission photons from an observed fluorophore image on the camera detector is unknown but can be represented as a random variable with a probability distribution including the phenomena of photon noise, bleaching and blinking. The Expected Maximum (EM) algorithm[41] can be applied to estimate the background mean and variance based on the intensity distribution of an averaged version of intensity. A Laplace of Gaussian (LoG) filter first provides particle recognition by template matching particle.[42] Then, a morphological filter is used to find the local intensity centroid for each candidate particle to be tracked. Each particle's pixel intensity distribution is then fitted by a two-dimensional Gaussian model to refine the estimate of intensity centroid to within sub-pixel precision, better than the standard optical resolution limit of our imaging system of 200-300 nm by a factor of ~10 (Fig. 4).[7,8,16,17]

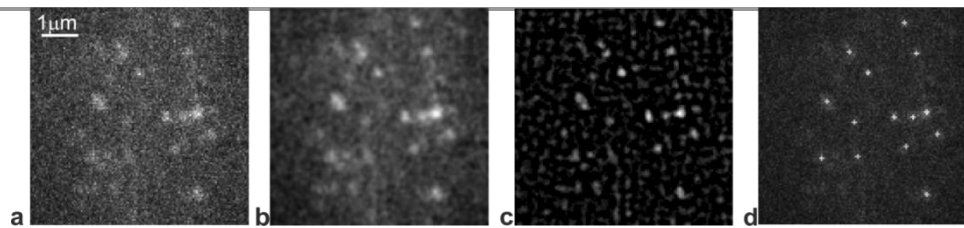

Fig. 4 Particle spot detection (a) Original fluorescence microscopy image with dynamical particle; (b) Smoothing filter image from template filter by Laplace of Gaussian (LoG); (c) Greyscale image based on successive geodesic morphological operations; (d) Final detection image with marker on each particle.

In a low SNR regime typical of live-cell single-molecule experiments not all fluorescently-labelled molecules may be detected. However, the detection likelihood can be characterized using stochastic simulated images with values of signal and noise typical of experimental data, with a detection probability of ~80% being typical for a single fluorescent protein of GFP under typical video-rate *in cellulo* imaging conditions from cell membrane studies of bacteria (Fig. 5a, red arrow trace), which improves dramatically with brightness of the observed fluorescent spot (for example, the presence of more GFP molecules in higher stoichiometry molecular complexes). Molecular mobility potentially affects detection likelihood however since it results in a blurring effect of the observed PSF fluorescent spot on the images, which typically can reduce the real detection probability by a further factor of ~2 (Fig. 5b, blue arrow trace).

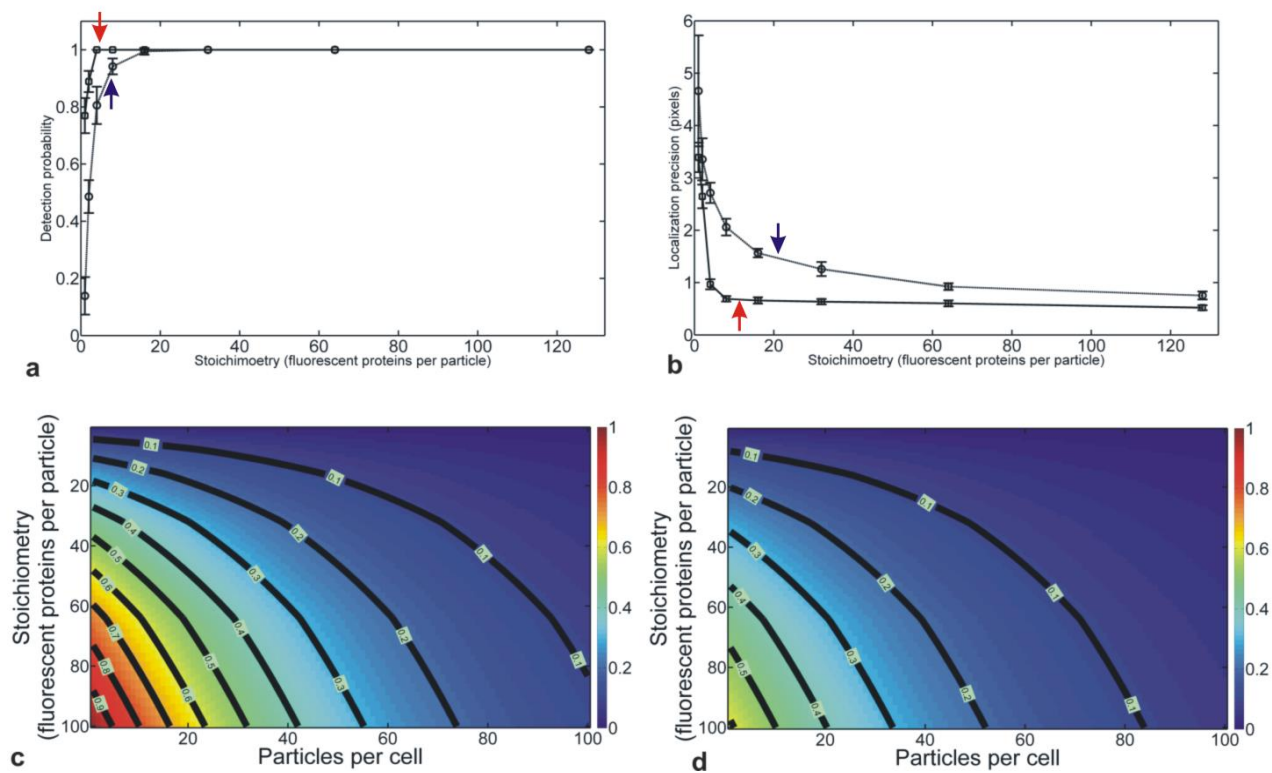

Fig. 5 Monte Carlo simulation to indicate the (a) detection probability and (b) localization precision as a function of stoichiometry (assuming a molecular complex is labelled here with typical GFP molecules observed using video-rate imaging typical of the studies used previously in my laboratory), with zero diffusion (red arrow trace) and a diffusion coefficient of 0.4μm$^2$/s (blue arrow trace), s.d. error bars. The probability of chance co-localization using an analytical Poisson model is indicated in (c) for zero diffusion, and (d) for a diffusion coefficient of 0.4μm$^2$/s, for the same imaging conditions as (a) and (b).

The spatial precision of pinpointing the intensity centroid of an observed fluorescent spot scales roughly with the reciprocal of the square-root of the number of photons sampled, simply due to Poisson sampling statistics. Realistic simulated images indicate that a lateral precision of 50-150 nm is typical for a single GFP molecule imaged at video-rate in the cell membrane of a living bacterial cell under our *in vivo*, video-rate conditions (compared with 15-20 nm lateral precision for many typical fixed-cell PALM studies), but as expected higher stoichiometry complexes would have improvements to this precision (Fig. 5b, red arrow trace). Again, molecular diffusion results in reductions to the observed localization precision due to the effect of image blurring (Fig. 5b, blue arrow trace).

Reliable detection of fluorescently-labelled single molecule also depends on their local density, namely the nearest neighbour separation distance; if this distance is comparable or less than the optical resolution limit as set by the PSF of the imaging system then there is a higher probability fluorophores will no longer be resolved individually as distinct molecules. The effect of increasing concentration of photoactive molecules on nearest-neighbour distance can be characterized analytically by using a Poisson nearest-neighbour model function[8,43] to generate maps for dependence of the likelihood of 'chance' co-localization (i.e. that two particles in a live cell will be separated randomly by less than the optical resolution limit) as a function of both the stoichiometry of diffusing particles (i.e. the number of fluorescent dye molecules per particle) and the number of diffusing particles per cell (Fig. 5c). Once again, non-zero molecular mobility has an effect here in reducing the probability of observing single distinct molecules by a factor of ~1.5 for typical diffusion rates observed in the cell membrane for protein complexes (Fig. 5d).

Once a candidate fluorescently-labelled molecule has been detected then the challenge lies in reliably linking together the same molecules in subsequent images to form a track. Single particle tracking (SPT) trajectories of fluorescently-labelled molecules in living cells are conventionally measured with respect to the global Cartesian camera detector coordinate system. However, in studies involving the investigation of membrane processes, fluorescent particles may be constrained to move over the non-planar cell membrane surface, so the conventional Cartesian tracking analysis may be inadequate for detailed studies of intracellular dynamics.[44,45] This is particularly relevant to observations made throughout different points of the cell cycle, since cell shape displays marked morphological changes. Conversely, in transforming the coordinate system of each two-dimensional diffusing particle in the membrane into a local cellular coordinate system, such that the coordinate plane in the immediate local vicinity of the particle is always parallel to its motion, the observed particle trajectory becomes independent of the translation and orientation of the cell.

Tracking of fluorescently-labelled membrane proteins involves fluorescent particle detection, nearest-neighbour linkage and fragment integration. A common problem is that some particle trajectories are only observed in fragments due to photophysics such as blinking or stochastic noise, so trajectories might not be linked due to missing image frames. In my laboratory, we have developed robust methods to link tracking segments into complete pseudo-three-dimensional trajectories at sub-pixel resolution by searching over minimum energy curves along the temporal axis using global combinatorial parameter optimization. We validate the performance of our algorithms using simulated data and apply it to experimental data obtained from live cells using a single-molecule fluorescence microscope, indicating a new, robust approach applicable to many different biological systems.

For SPT, traditional approaches depend on locally correlated information by linking particle feature points in images directly. Tracking fluorescent particles in a cell membrane is complicated because the trajectories may be interpreted based on different biological phenomena and particle diffusion models, such as Brownian or normal diffusion via a two-dimensional random walk, directed motion, confined diffusion, or anomalous or sub-diffusion.[46] Similarly, photophysics of the fluorescent reporter tag on the particle being tracked may add complications resulting in, for example, particle blinking manifest as truncated particle trajectories. Nearest-neighbour methods are simple to implement but very sensitive to noise.[35,47] Statistical methods, for example using probability density functions such as Multiple Hypothesis Trackers and NP-Hard methods have been applied with varying degrees of success.[40,48] Multiple Hypothesis Trackers (MHT) introduce probabilistic knowledge, but can be difficult to apply from matching a variable number of image feature points in global optimization both in space and time, while NP-Hard methods prohibit fast computation. Heuristic methods can be used to identify putative tracks from qualitative descriptions, but suffer the disadvantage of relying critically on the accuracy in the detection stage and easily fail when ambiguities occur.[48,49]

To overcome many of the shortcomings of existing tracking methods, my laboratory has developed a new automated multi-particle tracking algorithm based on minimal path optimization, similar to those used previously but extended in application to native cell membranes of living bacterial cells. After detecting candidate particles and linking image feature points frame-by-frame, some segmented trajectories are obtained initially. SPT trajectory data from time-lapse TIRF microscopy are combined from individual truncated tracks to create much larger trajectories using a pseudo-three-dimensional volume, and then the track in this pseudo-volume space from each moving particle is obtained using a combination of a minimal energy path approach mediated through a Fast Marching method,[50] a Dynamic Programming approach,[51] and the Linear Assignment Solution.[52]

To extract a particle trajectory, our method consists of three modules. Considering the particle intensity models from fluorescence microscopy, first a set of filters are applied to the image data to suppress noise and enhance the contrast of moving particles in each image frame. Then these particles are linked based on a nearest-neighbour principle between successive frames. However, a combination of high noise and complex particle movement often generates many truncated fragments of particle trajectories. The Grey Weighted Distance Transform (GWDT) method can group and complete such truncated tracks to form re-annealed trajectories using probabilistic criteria.[53] The final segmented pseudo-3D curve is linked by pairs of points from the tail of one trajectory to the head of another trajectory.

These methods are comparatively robust in dealing with ambiguities due to particle image fusion, missing detection events and appearance/disappearance of multiple targets in fluorescent particle tracking. The trajectories are then generated from the minimal energy path as defined by the solution of the time-dependent partial differential equation from the GWDT method. This approach offers a resilient tracking technique for the study of sub-cellular single-molecule dynamics in cell membranes.

Such localization and tracking methods may also be extended to new forms of so-called 'four-dimensional' microscopy that monitor fluorescently labelled molecules in three spatial dimensions as a function of the fourth dimension of time.[54-56] A simple but yet robust method is to introduce an asymmetry in the fluorescence emission optical pathway prior to imaging on a camera detector. In my laboratory we use a cylindrical lens component in tandem with the final imaging lens that projects the image onto an EMCCD detector via a dual-view filter that separates the image into two wavelength regimes for dual-colour imaging. The astigmatism induced deforms the image projected on the camera in a very specific way: if a point source is located in the microscope specimen focal plane then the image will be essentially identical to that of a non-astigmatic system, but when the source moves out of the focal plane a distinct deformation of the image occurs. The effect is that the symmetric image of a point source in the focal plane elongates in either the *x* or *y* direction depending on whether the fluorophore is moving closer or further away from the objective lens (Fig. 6). Using iterative unconstrained Gaussian fitting to these deformed PSF images allows the precise width in *x* and *y* to be determined, which then can be used in conjunction with prior calibration to determine the *z* position, typically to within a precision of 50-100 nm for single-molecule fluorescence imaging.

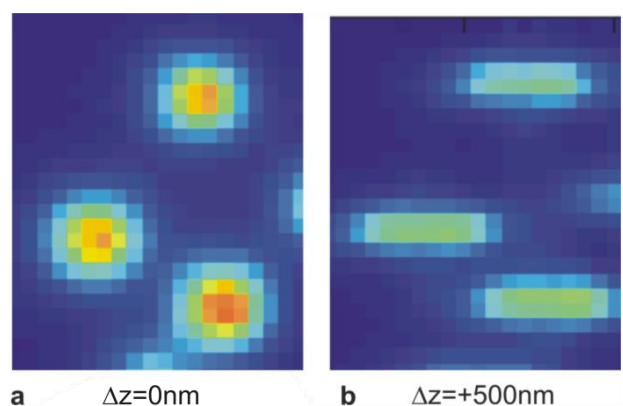

Fig. 6. Four-dimensional fluorescence microscopy (three spatial dimensions plus time) using astigmatism imaging (a) in the focal plane of fluorescent 200nm diameter beads, and (b) +500nm from the focal plane in *z*.

Putative molecular interaction may be quantified using dual-colour single-molecule fluorescence imaging. Here, the ability to robustly quantify if two molecules are definitively in the same region of space at the same time in the same living cell, namely that they are 'co-localized', is critical. In an ideal world one would potentially use Förster resonance energy transfer (FRET), however for live-cell applications at the single-molecule level this has proved particularly challenging technically due primarily to the poor photophysics of fluorescent protein FRET pairs used in monitoring the intercellular environment. However, significant information can still be obtained by using robust methods of co-localization analysis.

A method developed in my laboratory which has proved robust involves the use of fluorescence correlation. Firstly, to overcome the grainy quality of the raw images resulting mainly from the shot noise associated with low photon intensity and readout noise associated with the camera detector the graininess is smoothed out using a low-pass filter of a two-dimensional Gaussian kernel of width comparable to PSF of our imaging system (200-300 nm), which dramatically reduces noise associated at the level of individual camera pixels (Fig. 7).

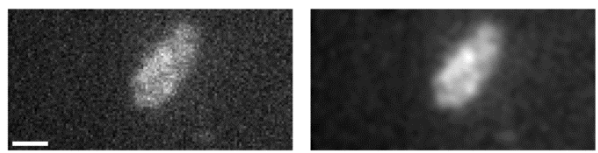

Fig. 7. Raw image of a fluorescent cell expressing fluorescent protein tagged OXPHOS complexes in the cytoplasmic membrane of *E. coli* (left panel) with low-pass two-dimensional Gaussian convolution (right panel), white 1 μm, strains developed in ref. 43.

To enable comparison of cell images from different colour channels used in dual-colour imaging the pixel intensity ranges are first normalised to be from zero to one in each channel. This step results in a loss of information on absolute intensity, but enables comparison of feature location in images with differing total intensities. To produce initial insight into spatial correlation, the channels of each (monochromatic) frame are combined into a single RGB image. As well as the intensity variation present in the monochromatic images, these images include the dimension of 'hue'. The channel correlation at each pixel can be identified by the hue resulting from mixing of the green and red components. While retaining all information of relative pixels values within each cell image, the combined colour images are difficult to interpret due to the multiple gradations in hue and intensity. A single quantitative metric for correlation is required. Alternative statistical measures of correlation exist such as Pearson's correlation coefficient and Spearman's rank correlation coefficient,[57] and with further techniques the precision of co-localization can be accurately calculated.[58] Our complementary approach has the beauty of computational simplicity and efficiency to implement and gives a very immediate indicator of the extent of co-localization across a whole population of individual cells.

The highest value pixels in each cell are first distinguished from the rest. The images can then be compared based on the proportion of overlap between the high value pixels. A 'binarization threshold' is set as the upper quartile pixel value within each cell image, resulting in binary images with 25% of the pixels in each cell having the high value. When cell images of each channel are overlaid, putative co-localization is identified through the location and total number of overlapping high-value pixels. The approach loses information on relative pixel intensity, but can produce a single and consistent metric on cell correlation.

Our fluorescence correlation technique is based on the realisation that in past and future frames, 'realistically redistributed' fluorescence is observable. So as well as comparing the channel images of one frame, channel images should be compared between frames from different times. The knowledge of what pseudo-independent (time displaced) co-localization looks like can then be used in the analysis of the co-localization in a single frame. This technique produces information on the motion of fluorescence from frame to frame which is used in interpretation of the correlation information.

For a given image sequence, either a colour channel is evaluated against itself (for example, green-green or red-red correlation) for information on realistic localisation and mobility, or one channel is evaluated against the other (green-red correlation) for information on molecular correlation. To explain the technique, a grid can be formed in which the rows and columns are associated with the sequence of frames of the two channels under analysis. For example, the rows (top to bottom) could be associated with the frames of the green channel and columns (left to right) the frames of the red (Fig. 8). The correlation of images associated with the row and column for each position in the grid is calculated. This means that a correlation measure is determined not only for images taken at the same time point, but also between frames displaced in time. The hue of each element signifies the colour channels that have been compared, and the brightness indicates the correlation value.

To infer a high likelihood of co-localization, values along the main diagonal are compared to the rest. If brighter than average, this indicates that the spatial distribution of intensity in both channels is correlated. If the intensity can be shown to be from the fluorescently labelled proteins, this indicates co-localization of the labelled molecules. There is total co-localization when one channel is compared with itself (as in Fig. 8 left and centre) where the diagonal elements have the maximum brightness. This is not necessarily the case when images of different channels are compared (as in Fig. 8 right).

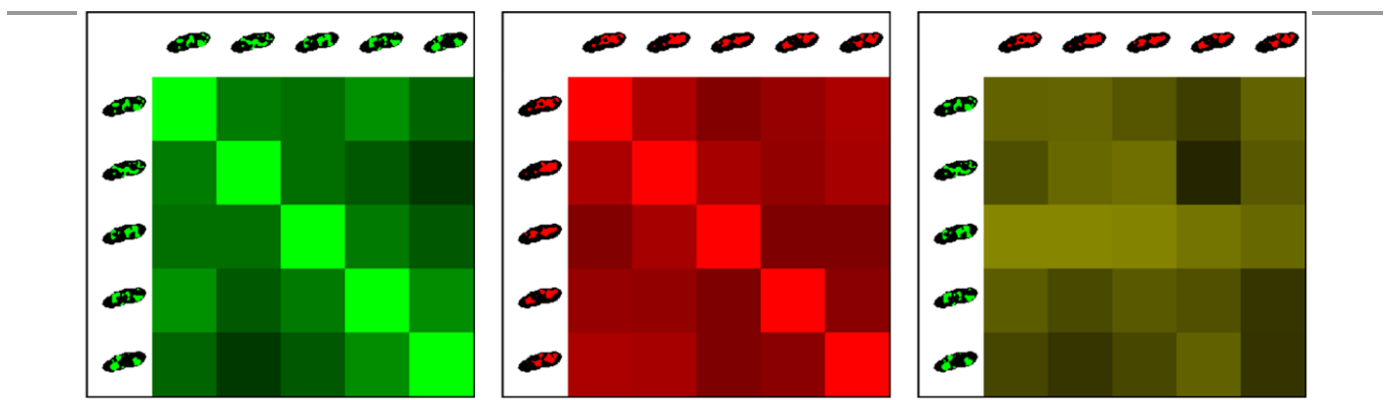

Fig. 8 Using fluorescence correlation to determine molecular co-localization. Arrays illustrating level of correlation between each image frame of five frame video sequences. (Left) Green-green correlation, (centre) red-red correlation, and (right) green-red correlation. Brighter colour corresponds to higher correlation. Cell strains developed in ref. 43.

## 3. Measuring molecular mobility

The mobility behavior of fluorescently-labelled single molecules in live cells may have several modes that deviate from simple Brownian motion, including directed diffusion, sub-diffusion also known as anomalous diffusion, as well as potential confinement effects. Diffusion analysis has until very recently been most popularly performed using relatively simple analysis of the mean-squared displacement (MSD) of tracked particles, but these heuristic measures rely on particle tracks of relatively large numbers of data points. However, due mainly to inferior photophysics in live cell studies only short tracks are typically obtained implying that stochastic effects become dominant. To counter this, my laboratory has developed a novel method called Bayesian Analysis to Ranking Diffusion (BARD) that uses propagator functions of diffusive processes directly to discriminate different modes that is capable of working on short fluorescent tracks.[46]

Brownian motion represents 'normal' diffusion characterized by a linear relation between time interval in the molecular trajectory and MSD. However, a tracked protein trajectory for which the MSD plateaus at large values of time interval indicates confinement suggesting that the protein is trapped by its local environment; such corrals may be important to forming nano-chambers for reactions thereby greatly enhancing the physical chemical efficiency. Directed diffusion has a typically parabolic MSD versus time interval trace, seen during active diffusive processes such as those of translocating molecular motors.[59,60] Anomalous or sub-diffusive behavior[61] is usually modeled as MSD proportional to time interval to the power of an exponent α where α is a coefficient between 0 and 1, indicative of percolation through the disordered media of the cell as well as a putative hopping motion between different confinement domains in the cell across corrals or interactions with specialized domains.[62-65]

Bayesian inference quantifies the present state of knowledge and refines it on the basis of new data. This posterior distribution incorporates any prior understanding on the set of parameters that comprise that model. Such priors embody our initial guess of the system, such as the expected order of magnitude or distribution of the parameters. The prior probability is independent of data of the system. The results of this inference are summarized by the most probable parameter values and their associated distributions, embodied in the posterior distribution of the parameter. After this stage, model comparison takes place in which diffusion models are ranked, conditioned by the observed data to assign a probability-based preference between the distinct models. If the model priors are flat (no particular a priori preference), the result of model evaluation is by simply ranking the marginal likelihood of each individual diffusion mode. This is also known as the evidence or the marginal likelihood and is given by integrating the data over the parameter space.

To implement the BARD diffusion analysis algorithm, firstly all the microscopic diffusion coefficients from the dataset are estimated from the initial gradient of the MSD trace of each molecular trajectory. Then the distribution of all microscopic diffusion coefficients is constructed and modelled by Gamma distribution. The shape parameters of this Gamma distribution are used to estimate the equivalent propagator function for each candidate diffusion mode, allowing a normalized marginal likelihood to be estimated for each individual track for each candidate diffusion mode, which can then be ranked and an inference thus made as to the most probabilistic diffusion mode to account for the individual trajectory data.

The use of the diffusion propagator functions in this way permits robust discrimination between Brownian, directed, confined and anomalous diffusion, even for relatively sparsely sampled data tracks as short as ~10 data points, relevant therefore to the truncated trajectories typically obtained from single-molecule live-cell fluorescence imaging. Separating molecular mobility characteristics into different categories offers enormous insight into several important physical chemistry questions concerning the living cell internal environment: how do proteins partition dynamically in different regions of the cell, how are signalling events linked to local sub-cellular architecture, how does the heterogeneous internal cell environment affect the mobility of motor proteins, and the extent to which interacting proteins rely upon random collisions or are part of putative confinement nano-reaction zones.[66,67]

## 4. Counting molecules in fluorescently-labelled complexes to quantify stoichiometry and turnover

If subunits within a molecular complex can be fluorescently labelled then the observed integrated fluorescence intensity from such a complex, when summed over all corresponding camera pixels within the diffraction-limited PSF fluorescent spot image, can be correlated to the total number of subunits present, namely the subunit stoichiometry of the complex, provided the brightness of a single fluorescent probe is known. In general this can be measured by utilizing the phenomenon of step-wise photobleaching of fluorophores, such that the size of the step in intensity between the light and dark states of a fluorophore is simply equivalent to the mean average brightness of that particular dye molecule. For complexes containing fewer than ~6 subunits these step-wise changes can be observed individually from a typical photobleach trace from a single molecular complex, and so the number of steps in the trace can be counted simply by eye, or by using some relatively trivial analysis routine, to indicate the subunit stoichiometry.[68]

However, for more challenging general cases of higher stoichiometry complexes, or oligomers or complexes, a more robust method is needed. The method developed for achieving this in my laboratory utilizes Fourier spectral analysis.[17] In essence, a pair-wise difference distribution is calculated for the whole of a single photobleach intensity versus time trace, obtained though continuously illuminating a single fluorescently-labelled molecular complex, and a power spectrum is then calculated for this pair-wise distribution of intensity values such that the fundamental peak in the power spectrum corresponds to the characteristic periodicity of the step-wise decrements in intensity during the raw photobleach trace.

Such raw steps are due to integer multiples of fluorophores undergoing step-wise photobleaching during a single sampling time window, therefore this characteristic periodicity is identical to the mean intensity $I_F$ of a single fluorophore during that photobleach. Poisson statistics indicates that the photoactive dwell time of a single fluorophore is exponentially distributed, implying that a general photobleach intensity versus time trace for several identical fluorophores can be fitted using a single exponential function of the form $I_0\exp(-t/t_b)$ where $t$ is time, $t_b$ is the characteristic photobleach decay time, and $I_0$ is the initial intensity given by the summed effects of all photoactive fluorophores. In the absence of any quenching effects or 'immature' dark fluorophores (such as fluorescent protein molecules that have not matured into their photoactive states following expression), $I_0/I_F$ is a measure of the number of proteins tagged with the fluorophore in the complex, in other words the molecular stoichiometry.

As an analytical tool this method is far more robust than more traditional approaches which rely on detection of individual step event in a noisy time series. Single-molecule experiments on living cells are rife with noise in general, with signals being sometimes only marginally above the level of the noise amplitude. Most molecular-scale events are manifest as some form of transient step signal in a noisy time series, the fluorescence intensity photobleach signature from a single fluorophore being one such example, and therefore the challenge becomes one of robust step-detection in a noisy data stream. Edge-preserving filters of raw data were originally employed – standard mean/spinal/polynomial-fitting filters perform badly in blurring distinct edges in a data stream. Median filters, or the Chung-Kennedy algorithm consisting of two adjacent running windows whose output was the mean from the window possessing the lowest statistical variance, preserve such edges; steps can then be detected as being probabilistically accepted or rejected on the basis of the change in the mean window output in light of the underlying noise (e.g. by calculating a corresponding the Student $t$-statistic for such a putative step change) between two adjacent windows run across a data stream time series, using a pre-defined threshold for acceptance. Variants of methods detecting steps from a noisy time series may be model dependent such that the probability for observing a step is history-dependent about earlier detection events, from so-called Markovian processes.[69]

However, such time-domain detection analysis algorithms are all sensitive to the level of detection threshold set; the acceptance threshold is often semi-arbitrary and subjective. Frequency-domain approaches, such as using Fourier spectral analysis as described above for photobleach traces but which have also been applied to other unrelated single-molecule studies such as the observed translocation of kinesin molecular motors on tubulin tracks, are potentially far less subjective since they utilize information obtained across the whole of a data trace as opposed to just a single putative step even in a data stream. The main disadvantage is the loss in time information for any specific individual step event in a given trace. Such analytical methods may also be employed for studies involving fluorescence recovery after photobleaching (FRAP) and fluorescence loss in photbleaching (FLIP) at single cell level and single molecule complex level to quantify the extent of dynamic molecular turnover.[17,18]

## 5. Rendering distributions of molecular behaviour

A recent improvement to objectifying single-molecule data is in how a distribution of single-molecule properties is actually rendered, e.g. step-sizes in terms of displacement of a tracked molecular motor, or the subunit stoichiometry measured from a molecular complex using step-wise photobleaching. Traditional approaches use some form of histogram to bin data across the observed distribution. However, it is clear that the size of a histogram bin and its position potentially lead to subjective bias. Such behaviour can lead to significant error in the general case of molecular heterogeneity – namely, that such distributions may be far from a unimodal, symmetrical Gaussian-type distribution, as is often mistakenly assumed to be the case, rather they may be far more complex, asymmetrical and often multimodal (for example, signifying the existence of metastable free energy states in the molecular probability distribution).

The development of analytical methods that instead use so-called kernel density estimation (KDE) has resulted in more objectivity in rendering molecular parameter distributions. With KDE, instead of data being pooled into semi-arbitrary histogram bins, the raw data are convolved using a detection sensitivity function, typically with a Gaussian function whose width is an estimate for the measurement error for that property in that particular experiment, and whose amplitude is then scaled such that the area under each unitary detection sensitivity function is equal to unity (reflecting a single detection event) – the sensitivity function thus indicates a realistic estimate for the actual sample distribution for a single event.

Histograms which contain too many bins potentially suggest there is more heterogeneity than there really is, whereas those which contain too few bins can hide heterogeneity. A KDE will generate the most objective distribution from any dataset for any single-molecule property (Fig. 9). This is important not simply on a qualitative level but rather the position of identified peaks in a distribution can be robustly quantified to identify distinct single-molecule states.

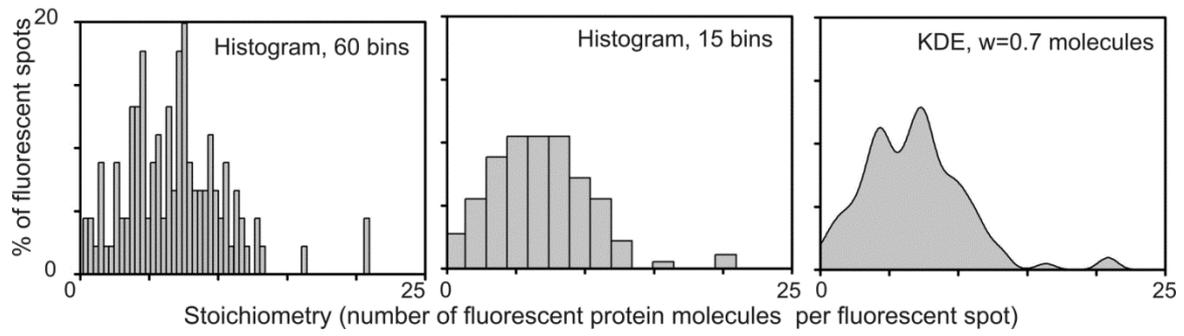

Fig. 9 Robust and objective rendering of the distribution of single-molecule parameters. Too small a histogram bin size (left) may suggest heterogeneity that is not really there, too few histogram bins (centre) may hide real heterogeneity. The most objective method to render a single-molecule parameter distribution is through kernel density estimate (centre), here shown on raw data obtained from ref. 18 on molecular stoichiometry of a component of the bacterial flagellar motor using a Gaussian convolution width of 0.7 molecules.

## Conclusions and Outlook

As microscope camera sensitivity, the photophysical properties of new fluorophores, and the methods of delivering fluorophores inside cells with specificity improve, single-molecule fluorescence imaging experiments in live cells will become increasingly more ambitious in terms of the practical aspects of imaging single molecular complexes in living cells[70]; truly multi-dimensional imaging using not just multiple colours[71] but multiple polarization states as well as simultaneous electrical and chemical measurements will most likely become increasingly more prevalent. Such complex datasets will ideally provide correlated and orthogonal information of molecular and cellular properties, necessitating yet further objective, analytical tools for extraction of molecular level information in a noisy environment. A natural extension of single-molecule research on single living cells is to rise to the cell population challenge – to perform single-molecule imaging of several cells simultaneously which potentially operate as a emergent multi-cellular, integrated structure, such as biofilms, tissues, even organs. An obvious challenge here is computational – having the ability to efficiently analyse molecular level tracking data from multiple cells simultaneously and, ideally to do so 'on-the-fly', namely in real-time such that the analysis is sufficiently fast to permit levels of feedback intervention to be applied to living sample. The logical basis to begin this challenge is to develop further robust analytical protocols to single-cell single-molecule data.

As the esteemed 19[th] century biologist Thomas Henry Huxley noted, the great tragedy of science is the slaying of a beautiful hypothesis by an ugly fact – the days of making qualitative judgements by eye for single-molecule biology research consistent with well-behaved 'beautiful' hypotheses have well and truly gone and the ascendance of the increasing ugly, complex, but very precisely known, single-molecule 'fact', is most certainly upon us.

## Acknowledgements

The author was supported from a Royal Society University Research Fellowship, the EPSRC, and the Biological Physical Sciences Institute (BPSI) at the University of York. Raw data were kindly donated by Quan Xue, Oliver Harriman and Erik Hedlund. Many thanks to Nick Jones for critical discussion. The author would particularly like to thank his students, postdocs and collaborators co-authored on many of the studies cited in this article, and to the growing collegiate fellowship of the BPSI at the University of York.